\documentclass[osajnl2,oneside,twocolumn,showpacs]{revtex4-1}
\usepackage{amsmath}
\usepackage{amssymb}
\usepackage{graphicx}

\begin{document}
\draft \title{Experimental Studies of the Mechanisms of Photomechanical Effects in a Nematic Liquid Crystal
Elastomer} \author{Nathan J.
Dawson$^{1,3 \dag}$, Mark G. Kuzyk$^{1}$, Jeremy Neal$^{2}$, Paul Luchette$^{2}$, and Peter Palffy-Muhoray$^{2}$}
\address{$^1$Department of Physics and Astronomy, Washington State University \\ Pullman, Washington  99164-2814 \\
$^2$Liquid Crystal Institute, Kent State University \\ Kent, OH 44242 \\
$^3$Currently with the Department of Physics and Astronomy, Youngstown State University \\ Youngstown, OH 44555 \\
$^\dag$Corresponding author: dawsphys@hotmail.com}
\date{\today}

\begin{abstract}

Azo-dye doped liquid crystal elastomers (LCE) are known to show a strong photomechanical response. We report on experiments that suggest that photothermal heating is the underlying mechanism in the surface-constrained geometry. In particular, we use optical interferometry to probe the length change of the material and direct temperature measurements to determine heating. LCEs with various dopants and optical density were used to study the individual mechanisms. In the high dye-doped limit, most of the light is absorbed near the entry surface, which causes a local strain from photothermal heating and a nonlocal strain from thermal diffusion. The results of our research on the microscopic mechanisms of the photomechanical response can be applied to designing photomechanical materials for actuating/sensing devices, the potential basis of smart structures.

\end{abstract}

 \maketitle
\section{Introduction}
\label{sec:introduc}

Alexander Graham Bell demonstrated that voice could be encoded on a beam of light, which when converted to an electrical signal, drove a speaker to reproduce the voice at a remote location.\cite{ja:bell00.01} Using a mechanical chopper to modulate a beam of light and focusing it on a variety of materials, he demonstrated that light could be directly converted to sound. As such, Bell's is the first known report of the demonstration of the photomechanical effect.

More recently, Uchino and coworkers demonstrated that photostrictive materials could be used to make a walking device that is powered by modulated light.\cite{uchin93.01,uchin90.01,uchin80.01} Subsequently, an all-optical position stabilizer was demonstrated in a dye-doped polymer fiber that was capable of keeping the $30$cm long fiber stable to within a tolerance of $3$nm.\cite{welke94.01} More significantly, this device combined all device classes,\cite{kuzyk06.06} which includes sensing, optical encoding of information, transmission, logic, and actuation. This device was later miniaturized into a polymer fiber interferometer, and shown to have multiple length states for a given input intensity (i.e. optical multi-stability).\cite{welke95.01} Later, such a device was used to demonstrate all-optical modulation, where one beam is used to change the length of the fiber interferometer, thus changing the transmittance of a second beam.\cite{welke96.01}

The last decade has shown increased interest in photomechanics. It has been reported that a reversible shape change can be created by photo-isomerization of nematic liquid crystal elastomers.\cite{corbe09.01,Finke01.01} Also, a highly doped liquid crystal elastomer (LCE) has been observed to swim on the surface of water in response to light.\cite{camac04.01} The mechanisms of the photomechanical effect were studied in a dye-doped fiber and were determined to be dominated by photothermal heating and molecular reorientation due to trans-cis isomerization leading to a bending effect and creating a polymer fiber cantilever.\cite{bian06.01,kuzyk06.06} The fact that long polymer optical fibers can be fabricated\cite{kuzyk91.01} with Bragg gratings\cite{chu05.01} and electrodes,\cite{welker98.01} adding photomechanical functionality may lead to a totally new class of devices.

The goal of this paper is to build an understanding of the mechanisms of the photomechanical effect in LCEs.\cite{NloSourceSmart} We attempt to decouple the photothermal heating and photo-isomerization mechanisms through a series of experiments. The results of our work may be important in applications such as large-scale parallel beam photomechanical optical devices (PODs).\cite{dawson10.01}

\section{Experiment}
\label{sec:experiment}

\begin{figure}[t!]
\includegraphics[scale=1]{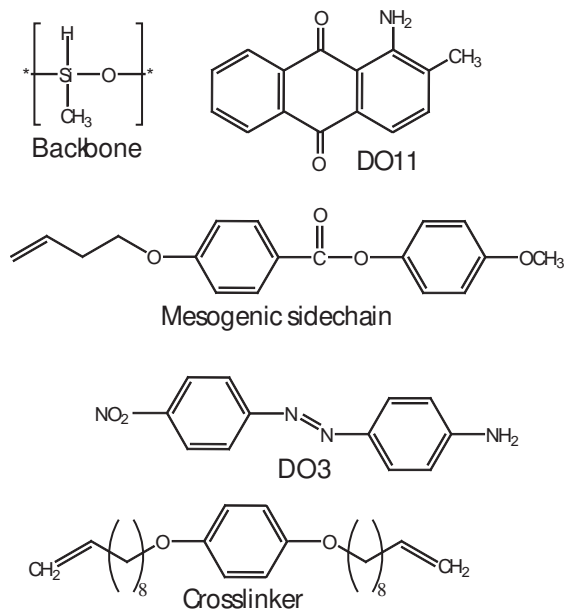}
\caption{The chemical structures of the silicon backbone, crosslinker, mesogenic sidechain, disperse orange 3 dopant chromophore, and disperse orange 11 dopant chromophore that are used to construct the dye-doped LCEs.}
\label{fig:chemmol}
\end{figure}

An LCE sample is pressed between two parallel glass substrates with the minimum force required to detect a ``wet spot" at each interface.  The inner surfaces of the glass substrates are partially coated with silver in the region away from the LCE. The LCE consists of a silicon backbone, crosslinker, and mesogenic sidechain whose structures are shown in Figure \ref{fig:chemmol}. The sidechain is connected to the backbone by the crosslinker. Every LCE used in the experiments presented in this paper has approximately $10$ silicon backbone segments per crosslinker. A light absorbing dye, photo-isomerizable dye or non-isomerizable dye is dissolved into an LCE to act as a light absorber that changes shape or transfers heat to the surrounding mesogens to yield a photo-induced strain.

\begin{figure}[b!]
\includegraphics[scale=1]{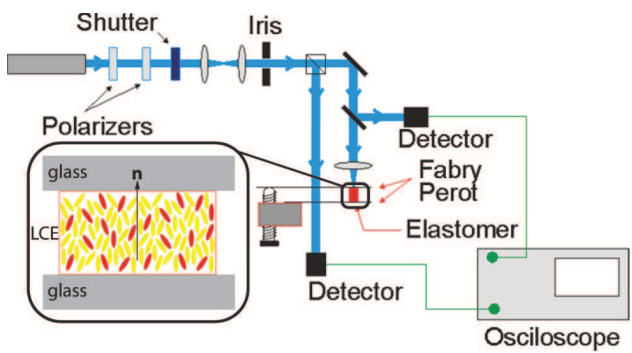}
\caption{(Color online) Schematic diagram of the experiment used to illuminate an LCE and to measure it's length change using the interferogram pattern of the reference beam.}
\label{fig:figure3}
\end{figure}

Prior to placing the LCE between two glass slides to make a sample, it is prepared from a larger specimen by stretching to induce alignment. A small section of this larger LCE is cut from its center to obtain a material with uniform order parameter. During typical experiments, the change in length of the LCE is small compared with the length change corresponding to the transition from isotropic to aligned, thus, the equilibrium photomechanical length change is a linear function of the laser power.

The sample is tacky, so it sticks to the substrate and cannot slip along the interface in the presence of small transverse forces, as should be the case for small photomechanical strains.  Furthermore, because the sample supports the weight of the upper substrate, the photomechanical response is measured for constant stress in the uniaxial direction and under the boundary constraint of no shear at the glass-LCE interface.

The director of the LCE is perpendicular to the glass substrates. One of the glass plates is mounted with a fixed support that acts as a pivot point and two thumb screws that define two orthogonal tilt axes. Thus, one of the mirrors can be adjusted to make it parallel to the other one. Figure \ref{fig:figure3} shows a schematic diagram of the experiment (only one of the adjustable supports is shown).

The laser is split into two paths. One of the beams passes through the LCE and the second beam passes through the two silvered parts of the glass plates, which act as a Fabry-Perot interferometer. Changes in the length of the LCE results in a change in the output of the interferometer. The light intensity that is transmitted through the interferometer and read by the detector can be used to determine the change in length of the LCE. A small portion of the light incident on the LCE is reflected and read by a detector. The laser beam is instantaneously turned on at time $t=0$, and the waveforms are recorded by an oscilloscope. After a long time, the laser is turned off, and the transient behavior is again recorded until the system reaches equilibrium.

The procedure above is repeated for a range of intensities, keeping all other parameters fixed. Enough time is permitted to elapse between experiments to prevent bias in the orientational order parameter.

The optical path length was found using optical absorption spectroscopy. First, a thinly-sliced LCE is pressed along the director between two glass slides until the sample thickness is approximately $10\,\mu$m. The absorption spectrum is then measured using a short enough pulse of white light to prevent photo-isomerization. Subsequently, the Beer-Lambert coefficient, $\mu$, is calculated for the pressed sample assuming an isotropic phase using
\begin{equation}
\mu = \frac{1}{l}\ln \frac{I}{I_0} .
\label{eq:BLcoeff}
\end{equation}
where $l$ is the thickness, $I$ is the transmitted intensity, and $I_0$ is the incident intensity. For a high concentration DO3-doped LCE probed at $488\,$nm in the isotropic phase, $\mu \approx 1\,\mu\mbox{m}^{-1}$.  Assuming that the order parameter, $Q$, vanishes in the highly-strained sample ($Q=0$), and estimating that $Q\approx 0.7$ under ordinary experimental conditions, we can use the definition of the order parameter, $Q = \left\langle 3\cos^2 \theta - 1\right\rangle / 2$, to get
\begin{equation}
\mu\left(Q\right) = \mu\left(Q=0\right) \frac{2}{3}\left(1-Q\right) .
\label{eq:BLcoeffQ}
\end{equation}
Thus, $\mu\left(Q=0.7\right)\approx 0.2 \,\mu\mbox{m}^{-1}$ for the high-concentration sample.

\section{Results}
\label{sec:eresults}

The goal of these experiments is to test the mechanisms of length change as a function of time for various laser powers. For small changes, the change of the order parameter is proportional to both the temperature change, $\Delta T$, and the trans isomer population fraction, $\Delta N$, or $\Delta Q_N \propto \Delta N$ and $\Delta Q_T \propto \Delta T$. Here, $\Delta Q_N$ is the change in the order parameter due to a change in the population fraction of trans isomers and $\Delta Q_T$ is the change in the order parameter from a temperature change. Note that the change in the order parameter is given by a sum over each contribution, $\Delta Q = \Delta Q_N +\Delta Q_T$. Due to the coupling between the mesogens and the elastomer backbone, a change in the order parameter will yield a change in length, $\Delta L$, i.e. $\Delta Q \propto \Delta L$. Therefore, under small length changes, these experiments decouple the mechanisms of photomechanical length changes.

\begin{figure}[t!]
\includegraphics[scale=1]{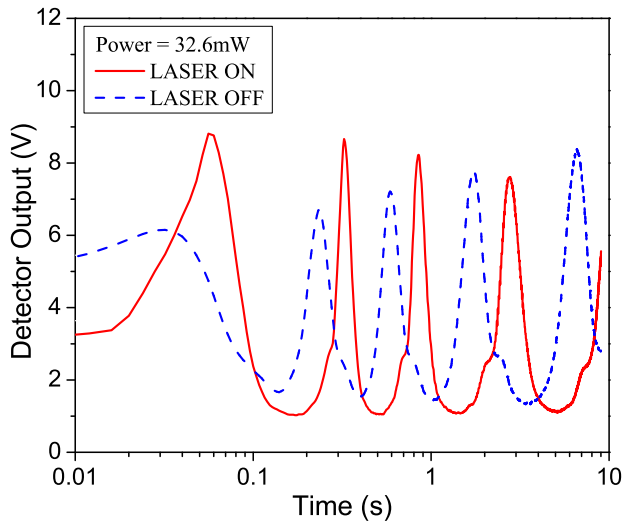}
\caption{(Color online) The measured transmitted probe beam intensity through the integrated Fabry-Perot interferometer, where the LCE is illuminated uniformly across its surface with a pump power of $32.6\,$mW.}
\label{fig:326ONOFF}
\end{figure}

Two sets of experiments are used to characterize the sample. First the photomechanical length change is measured with the optical interferometer. The length was found to decrease under illumination by tracking the center fringe placement of the probe beam. Then, a temperature probe is used to directly measure photothermal heating, as follows.

In a typical experiment, the LCE's length changes by several laser wavelengths. To simplify the interpretation of the transmittance of the probe beam through the interferometer, only the peak intensities were used to determine the separation between the substrates that define the mirrors. We note that there are small systematic errors introduced from the fact that the mirrors do not remain parallel for large length changes, as is evident from Figure \ref{fig:326ONOFF}, where the peak heights change as the plates lose alignment. The same drift is observed during pumping and relaxation and appears to be reversible, as would be expected.

\begin{figure}[t!]
\includegraphics[scale=1]{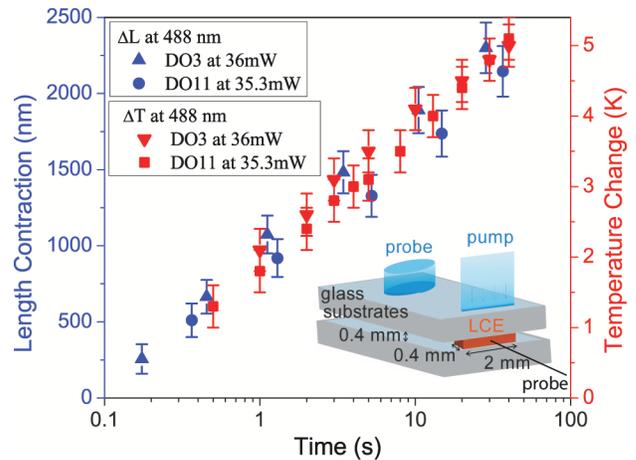}
\caption{(Color online) The degree of length contraction and temperature increase as a function of time for a DO3- and DO11-doped LCE. Inset shows a diagram of the sensor placement in the LCE.}
\label{fig:templength}
\end{figure}

\begin{figure}[b!]
\includegraphics[scale=1]{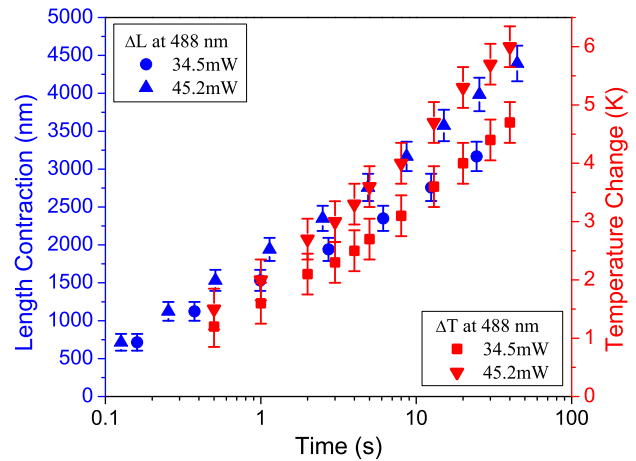}
\caption{(Color online) The degree of length contraction and temperature change as a function of time for a $0.02\%$ by weight DO3-doped LCE at two different laser powers.}
\label{fig:DO3LCE002bywt}
\end{figure}

A temperature probe that is tapered to a point relative to the length scale of the elastomer is inserted in the center of an LCE, well beyond the illuminated region, thus recording the average temperature of the sample as a function of time. The sensor is much thinner than the sample as shown in the inset of Figure \ref{fig:templength}. The length change is also measured in the sample as a function of time prior to using the optical probing technique described above. Length contraction and temperature change for a $0.1\%$ by weight DO3 dye-doped LCE subject to a $36\,$mW power laser is also shown in Figure \ref{fig:templength}. Note that an increase in temperature causes a decrease in orientational order, thereby inducing a negative strain. The observed correlation between the two suggests that the thermal mechanism is responsible. This data is consistent with the view that the laser light is absorbed near the surface of the sample, and the length change of the sample results from a temperature increase of the rest of the sample due to the flow of heat from the illuminated part of the LCE.

\begin{figure}[t!]
\includegraphics[scale=1]{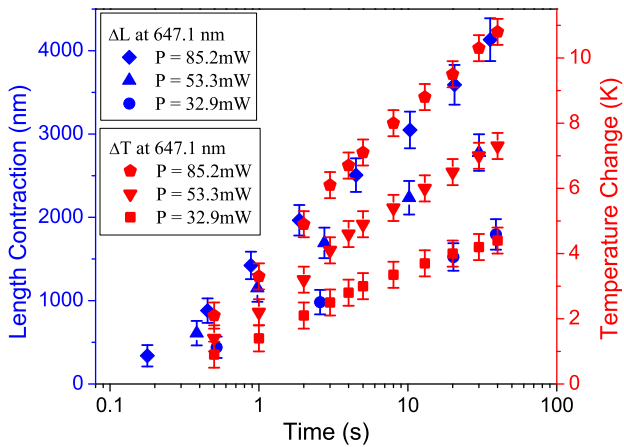}
\caption{(Color online) The length contraction and temperature change as a function of time for a $0.1\%$ by weight DO3-doped LCE over various laser powers at a wavelength of $647.1\,$nm.}
\label{fig:DO3HC647}
\end{figure}

A non-isomerizable dye, disperse orange 11 (DO11), was dissolved into a LCE as a control in which photo-isomerization-induced length changes are absent. The strategy was to check for deviations in the temperature change and length contraction data over time as an indicator of the photo-isomerization mechanism. Figure \ref{fig:templength} shows the length contraction and temperature change as a function of time for a high concentration DO11 dye-doped LCE. Not only is the rate of temperature change remarkably similar to the rate of length contraction as for DO3-doped LCEs, but the magnitudes for these two experiments are also equivalent within experimental uncertainty. This shows that two LCEs with approximately the same initial length and orientational order, one doped with photo-isomerizable chromophore and with a non-isomerizable chromophore, have the same rate of length change.

The length change due to photo-isomerization is negligible for a LCE with high DO3 concentrations due to the low penetration depth of light.  A lowered concentration increases the penetration depth of the pump light, but decreases the photo-isomerization induced strain due to the lower concentration of chromophores.  The lower number of trans isomers decreases the number of trans isomers that can be converted to cis isomers, which are responsible for interfering with nematic ordering forces, and thus decreasing the LCE's long-range order and also its length. The length and temperature changes of a LCE doped with DO3 at a concentration of $0.02\%$ by weight are shown in Figure \ref{fig:DO3LCE002bywt}. This confirms that the contribution of photo-isomerization is negligible even at low concentrations.

The magnitude of length contraction for the experiment shown in Figure \ref{fig:DO3LCE002bywt} is larger than the magnitude of length contraction for the experiments shown in Figure \ref{fig:templength}. This is due solely to the difference in initial order parameter of the two LCEs. We have found that the upper glass substrate, while lightweight, causes a slow decrease in the order parameter of the LCE due to the resulting uniaxial stress. At a lower initial order parameter, the photomechanical response is larger due to the LCE being closer to the nematic-isotropic phase transition. Thus, the photomechanical response of a sample can become larger over time.

\begin{figure}[t!]
\includegraphics[scale=1]{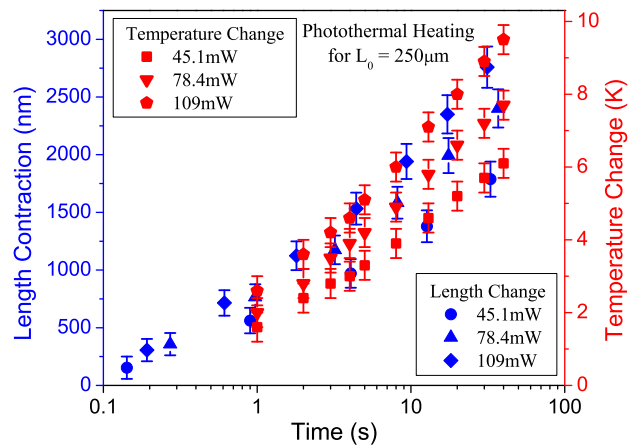}
\caption{(Color online) The temperature dependence of length contraction as a function of time for a high concentration DO3-doped LCE with initial length $L_0 \approx 250\,\mu$m.}
\label{fig:PTHeating}
\end{figure}

It is possible that photo-isomerization-induced length contraction was absent because the concentration of photonematogens was sub-critical for affecting the long-range order. Therefore, we tested a high concentration ($0.1\%$ by wt.) DO3-doped LCE that is illuminated by light that is far from the trans absorption peak of DO3. At $647.1\,$nm, the penetration depth is much greater than at the $488\,$nm wavelength light used above because $647.1\,$nm is far off-resonance in DO3. Figure \ref{fig:DO3HC647} shows the temperature change and length contraction as a function of time over a range of light intensities using the $647.1\,$nm pump.

The contribution from the photo-isomerization mechanism in the off-peak absorption experiments is found to be negligible within experimental uncertainty. This is once again based on the correlation between the temperature change and length change as shown in Figure \ref{fig:DO3HC647}.

To fully prevent photo-isomerization, we performed a control experiment in which a laser source is prevented from entering the LCE. To do so, a highly absorbing thin film is placed between the glass substrate and LCE. The laser is absorbed by the thin film, which results in heating of the LCE surface. Results from the pure photothermal heating experiment of a $250\, \mu$m thick and high concentration DO3-doped LCE that is illuminated at multiple intensities is shown in Figure \ref{fig:PTHeating}. The rate of heating and length change are equal within experimental uncertainty and exhibit the same trend on a logarithmic time scale. Since the rate of photothermal heating and length change are completely correlated in all experiments, we conclude that the photo-isomerization mechanism is negligible in our sample geometry.

\section{Empirical Model}
\label{sec:empirimodel}

The experimental results strongly suggest that photo-induced strain can be explained purely by a temperature increase due to photothermal heating. The absorption of light through the sample can be modeled using the Beer-Lambert law,
\begin{equation}
I = I_0 e^{-\mu z} ,
\label{eq:bllawLCE}
\end{equation}
where $z$ is the position in the LCE along the direction of the propagation of light, $I$ is the intensity at that position, $I_0$ is the incident intensity, and $\mu$ is the absorption coefficient, which is related to the imaginary part of the dielectric permittivity. The heat generated in the material due to the light source, $H_s$, is then given by
\begin{equation}
H_s = - \alpha \frac{d I}{d z} ,
\label{eq:intheatgenLCE}
\end{equation}
where $\alpha$ is a parameter that depends on the concentration and relates the intensity gradient to the amount of heat generated.

The heat diffuses through the LCE and exits at the glass/LCE interface boundaries via contact conductance. The heat in the glass diffuses through the outer glass interface to the remainder of the apparatus by convection. The apparatus geometry causes the time-dependence of the temperature to be more complex than one would expect from the idealized sample geometry shown in the insets of Figures \ref{fig:chemmol} and \ref{fig:templength}.

\begin{figure}[t!]
\includegraphics[scale=1]{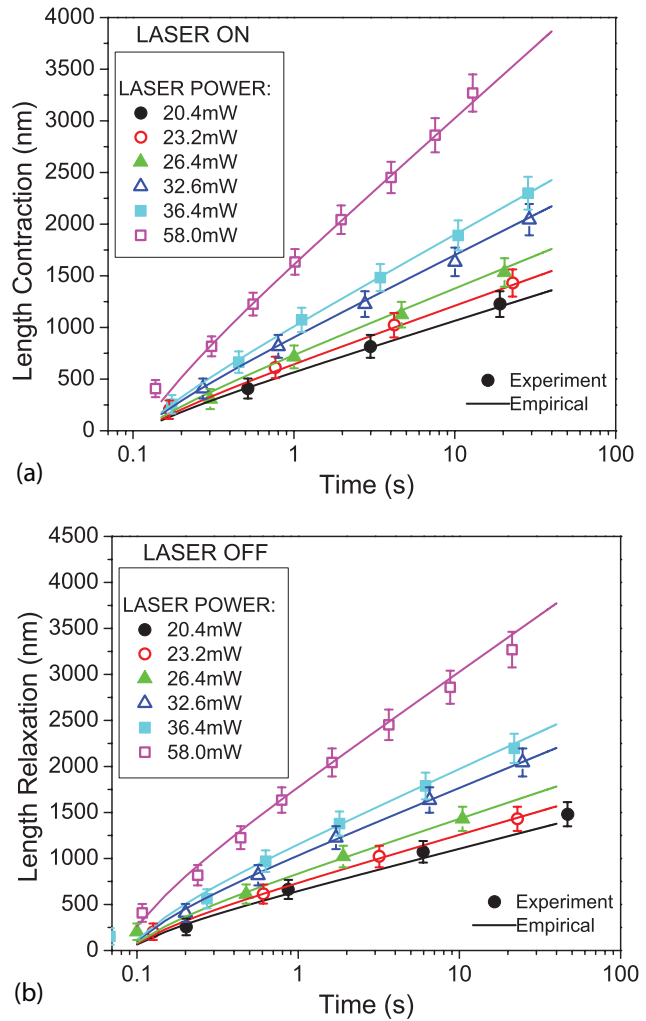}
\caption{(Color online) (a) Laser on: empirical fit with $k = 10.43 \pm 0.48\,$nm/mW, $\gamma = 14.88 \pm 2.58 \, \mbox{s}^{-1}$, and $\phi = -0.63 \pm 0.75$. (b) Laser off: empirical fit with $k = 9.59 \pm 0.36\,$nm/mW, $\gamma = 28.59 \pm 5.58 \,\mbox{s}^{-1}$, and $\phi = -1.46 \pm 1.47$.}
\label{fig:empirical}
\end{figure}

In order to simplify the characterization of the photomechanical effect in a material, it would be useful to have an empirical model that describes a broad range of materials and that depends on a minimal number of parameters that each characterizes a distinct property of the system.  Under our experimental conditions, the change in length as a function of time can be approximated by a logarithmic curve for  $t>1\,$s - when a steady state between the LCE and its surrounding environment is reached. Figure \ref{fig:empirical} shows a logarithmic plot of the experimental data and theoretical fits for the illuminated and dark relaxation processes.

The logarithmic model is modified with an empirically determined time offset, $\phi$, of the form,
\begin{equation}
    \Delta L = k P \ln (\gamma t + \phi) ,
\label{eq:eqno36}
\end{equation}
where $P$ is the incident power, $k$ is a constant that represents the magnitude of the length change, $\gamma$ is a factor that governs the rate of length change per unit of power, and $\phi$ is the time offset. It should be noted that $\gamma$ embodies several parameters, such as the LCE and glass thermal diffusivity, the contact conductance coefficient at the LCE-glass interface, and the convection coefficient at both the LCE- and glass-air interfaces. A change in any one of these properties will affect the rate of length change.  $k$, on the other hand, is related to the photomechanical strength, which for photothermal heating, is related to the optical density of the material and the coefficient of thermal expansion

The model yields a good approximation to the data, as shown in the top part of Figure \ref{fig:empirical} for an LCE illuminated with a $488\,$nm laser, as well as the relaxation process for the dark LCE, which is shown in the bottom of Figure \ref{fig:empirical}.  As such, it appears to be a suitable analytical model for the photomechanical effect.

\section{Discussion}
\label{sec:discuss}

The experimental results are consistent with the decoupled approximation because most of the energy is absorbed near the surface of the sample, acting as a source of heat that propagates into the bulk of the sample and leads to a thermal response. For small changes in length, the length change is proportional to the change in the orientational order of the mesogens. This observed length change is a complicated function of time that is inconsistent with a localized isomerization mechanism. Thus, the thermal mechanism is found to be dominant and decoupled from photo-isomerization based on both (1) a series of experiments with several control experiments that are insensitive to photo-isomerization as well as (2) the fact that the dynamics of the process depend on the LCE geometry and thermal properties of the experimental apparatus, not on the intrinsic properties of the LCE itself.

Due to the complexity of the system's thermal properties, the time evolution of the temperature profile can only be solved numerically. However, we propose a simple empirical model that with only two adjustable parameters can describe the dynamical behavior over a large range of times and intensities. Such a model would be useful for characterizing the photomechanical response of any material in which the underlying mechanisms originate in temperature changes that influence the order parameter when the surrounding material acts as a thermal reservoir.

We see our work as the first step in characterizing materials for the purpose of both making better materials, and for using this information for the design of smart photomechanical systems. This work was partially motivated by the results of dynamical studies of PODs in response to complex waveforms.\cite{dawson10.01} That research demonstrated that one could cascade a series of PODs where the output of one device drives the next device. Finally, one can imagine extending the Fabry-Perot cavity around the LCE to impart it with a nonlinear response and hysteresis. Interconnecting such nonlinear units together would lead to a network that could be programmed for ultra smart behavior.\cite{kuzyk06.06}

\section{Conclusion}
\label{sec:conclusionmech}

Using a series of experiments that each independently isolate a single variable, we have found that a dye-doped LCE's photomechanical response in a POD geometry is dominated by photothermal heating. We find that independent of dye concentration or pump intensity, the length change is well correlated with the measured temperature change.

In a set of control experiments, the photo-isomerization mechanism is eliminated by (1) using dyes that do not have isomers, and (2) blocking the sample with an absorbing film that prevents light from entering the sample but causes heating at the surface. All these control experiments show a correlation between length change and temperature change that matches the behavior of samples in which light propagates in regions of photo-isomerizable material. This set of experimental results provides strong evidence that photothermal heating is the dominant mechanism.

In future work, we plan to numerically model the photothermal heating mechanism as an independent test of our findings.  Such models can be used to design materials for smart photomechanical systems.

{\bf Acknowledgements:} MGK and PPM thank the National Science Foundation (ECCS-0756936)
for generously supporting this work. NJD and MGK thank the Air Force Office of Scientific Research (FA9550-10-1-0286) for their generous support.

\end{document}